\RequirePackage{fix-cm}
\documentclass[smallextended]{svjour3}       
\smartqed  
\usepackage{graphicx}
\newcommand{\ANMI}{A.1}
\newcommand{\ADemo}{A.2}
\newcommand{\AIntervention}{A.3}
\newcommand{\ACalibration}{A.4}
\newcommand{\AParameter}{A.5}
\newcommand{\AResults}{A.6}
\newcommand{\TDemo}{A.2.1}
\newcommand{\TInt}{A.3.1}
\newcommand{\TCal}{A.4.1}
\newcommand{\TPar}{A.5.1-A.5.7}
\newcommand{\TRes}{A.6.1}
\usepackage{longtable}
\usepackage{pdflscape}
\usepackage{multirow}

\begin{document}

\title{Use of mathematical modelling to assess respiratory syncytial virus epidemiology and interventions\thanks{The author is an employee of Merck Sharp \& Dohme Corp., a subsidiary of Merck \& Co., Inc., Kenilworth, NJ, USA, which supported this work. Funding sources were not involved in study design; collection, analysis, and interpretation of data; writing of the repot; or in the decision to submit for publication.}
}
\subtitle{A literature review}


\author{John C. Lang}


\institute{J.C. Lang \at
              Merck \& Co., Inc. \\
              Center for Observational and Real-World Evidence (CORE),
              Kennilworth, NJ, USA
              Tel.: +1-215-652-5978
              \email{john.lang@merck.com}
}

\date{Received: date / Accepted: date}

\maketitle

\begin{abstract}

Respiratory syncytial virus (RSV) is a leading cause of acute lower respiratory tract infection worldwide, resulting in approximately sixty thousand annual hospitalizations of $<$~5-year-olds in the United States alone and three million annual hospitalizations globally. The development of over 40 vaccines and immunoprophylactic interventions targeting RSV has the potential to significantly reduce the disease burden from RSV infection in the near future. In the context of RSV, a highly contagious pathogen, dynamic transmission models (DTMs) are valuable tools in the evaluation and comparison of the effectiveness of different interventions. 
This review, the first of its kind for RSV DTMs, provides a valuable foundation for future modelling efforts and highlights important gaps in our understanding of RSV epidemics. 
Specifically, we have searched the literature using Web of Science, Scopus, Embase, and PubMed to identify all published manuscripts reporting the development of DTMs focused on the population transmission of RSV. We reviewed the resulting studies and summarized the structure, parameterization, and results of the models developed therein. 
We anticipate that future RSV DTMs, combined with cost-effectiveness evaluations, will play a significant role in shaping decision making in the development and implementation of intervention programs.

\end{abstract}

\section{Introduction}
\label{S:Intro}

Respiratory syncytial virus (RSV), a highly contagious disease, has increasingly been recognized as a leading cause of acute lower respiratory tract infection worldwide \cite{NairEtAl2010,NairEtAl2013}. The overwhelming majority of individuals are infected by their second year of life \cite{GlezenEtAl1986,HendersonEtAl1979}. Severe disease is most common in young infants ($< 6$-month-olds), with incidence decreasing rapidly with age \cite{HallEtAl2013,HallEtAl2009,RhaEtAl2020}. Globally, it is estimated that RSV is responsible for approximately three million hospitalizations annually in children ($< 5$-year-olds) \cite{NairEtAl2010,NairEtAl2013}. Lifelong reinfections with RSV are common and, although healthy older children ($5-17$-year-olds) and adults ($18-64$-year-olds) are less likely to develop severe disease \cite{HallEtAl2001}, severe disease is more common in older adults ($\geq65$-year-olds), institutionalized individuals, and immunocompromised individuals \cite{FalseyEtAl2005,WidmerEtAl2014,WidmerEtAl2012}. Respiratory syncytial virus epidemics exhibit rich dynamics that vary geographically and climatically; both annual peaks, and biennial alternating high- and low-peaks, have been observed in RSV epidemics \cite{BloomEtAl2013,LiEtAl2019}.

At present there is only one immunoprophylaxis, the monoclonal antibody palivizumab, that is recommended for the prevention of RSV disease; however, due to its high expense and limited effectiveness, recommendations are generally limited to high-risk patients, i.e., very premature infants, infants with chronic lung disease (CLD), or infants with congenital heart disease (CHD) \cite{BradyEtAl2014,GutfraindEtAl2015}. Nevertheless, development of RSV immunoprophylactic interventions are proceeding rapidly, with over 40 RSV vaccines or immunoprophylactic interventions currently under development \cite{HigginsEtAl2016,PATH2020}.

Mathematical models play an important role in many aspects of epidemiological research \cite{ChubbJacobsen2010}. For example, the US Centers for Disease Control and Prevention has recently developed a static model that can be applied to evaluate the number of medically attended RSV infections subject to various interventions \cite{RainischEtAl2020}. Whereas static models are effective at estimating the direct effects of immunoprophylactic interventions, they are ill-suited to the study of indirect effects or herd immunity effects, which are frequently significant for infectious diseases \cite{PitmanEtAl2012}. Thus, in anticipation of the availability of multiple immunoprophylactic options for RSV there has been increasing interest in the development of dynamic transmission models (DTMs) that are fully capable of representing complex interactions between virus, environment, population, and immunoprophylactic interventions. As with static models, DTMs can be integrated into cost effectiveness analyses to aid public policy decision making with respect to the control of RSV.

The principal aim of this literature review is to provide an overview of RSV DTMs as a resource for future RSV dynamic modelling efforts. We proceed in four parts. First, we outline RSV DTM structures. Second, we summarize data sources used for model calibration and common parameter values determined through model parameterization. Third, we present the main findings of RSV modelling papers. Finally, we identify key areas for future modelling research and discuss how mathematical modelling can contribute to public health decision making.

\section{Search strategy and results}
\label{S:Search}

\subsection{Search strategy and selection criteria}
\label{S:SearchCriteria}

Studies for this review were identified through searches of Web of Science \cite{WoS2020}, Embase \cite{Elsevier2020b}, Scopus \cite{Elsevier2020a}, and PubMed \cite{NCBI2020}, by use of terms (a) “respiratory syncytial virus”, “human respiratory syncytial virus”, “rsv”, or “hrsv”, and (b) “mathematical model”, “dynamic transmission model”, “dynamic model”, “transmission model”, “epidemic model”, “compartment model”, or “compartmental model”. For Embase and PubMed searches we add corresponding Emtree (“respiratory syncytial virus” or “human respiratory syncytial virus”, and “mathematical model” or “dynamic transmission model” or “dynamic model” or “compartment model” or “compartmental model”) and MeSH (“respiratory syncytial virus, human” or “respiratory syncytial viruses”, and “models, theoretical”) search terms. Search terms were applied to all fields, all dates were included, no language restrictions were applied, and only published manuscripts were included.
 
\begin{figure*}
  \includegraphics[width=1\textwidth]{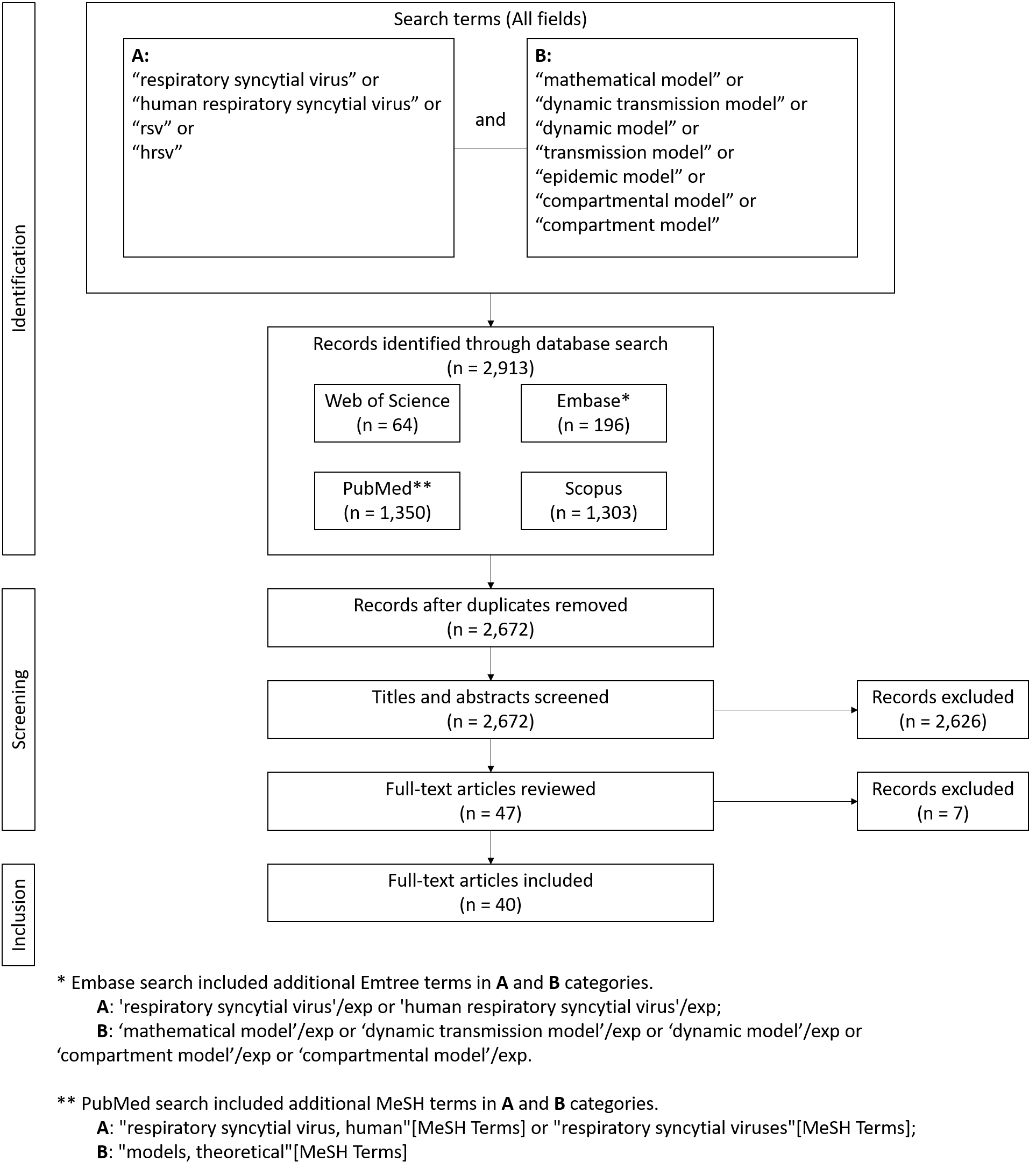}
\caption{Search strategy for idenitfying RSV DTMs}
\label{figure:Search}       
\end{figure*}

Duplicates, which were determined by exact match of title, authors, and year of publication, were removed. Subsequently, titles and abstracts were reviewed and a priori inclusion/exclusion criteria were applied. Inclusion criteria are manuscripts that have been published in peer reviewed journals and present a human epidemiologic RSV DTM (e.g., animal and immunologic models are not included). Multi-pathogen DTMs (e.g., a DTM modelling RSV and influenza concurrently) and ensemble models are excluded, as are manuscripts whose primary purpose is other than dynamic transmission modelling of RSV (e.g., a manuscript whose primary focus is the analysis of an abstract DTM in a general context, for which an RSV DTM is given as an example in passing). Full-text articles were retrieved for all manuscripts identified in title and abstract screening procedure. All full-text articles were reviewed in full; inclusion/exclusion criteria were re-applied during review of full-text articles. The sole author of this study performed all search steps in duplicate using Endnote~X9 reference management software. The search strategy is summarized in Figure~\ref{figure:Search}. A summary of data abstracted is given in Table~\ref{table:Search}. Data abstraction and verification were performed manually and in duplicate by the sole author of this manuscript; no data abstraction software was used. 

\begin{table}
\caption{Summary of data abstracted}
\label{table:Search}       
\begin{tabular}{p{2.5cm}p{5cm}p{3cm}}
\hline\noalign{\smallskip}
\textbf{Data abstracted} & \textbf{Description} & \textbf{Summary table(s)}  \\
\noalign{\smallskip}\hline\hline\noalign{\smallskip}

\noalign{\smallskip}
Disease state structure & Record the disease state structure used in the RSV DTM, e.g., $SIR$, $SIRS$, $SEIRS$, etc. & Table~\ref{table:Summary} \\

\noalign{\smallskip}
Modelling approach & Record the mathematical modelling approach used in the implementation of the RSV DTM, e.g., ordinary differential equation (ODE), stochastic differential equation (SDE), or agent-based model (ABM), etc. & Table~\ref{table:Summary} \\

\noalign{\smallskip}
Demographic model & Record whether a demographic model is present & Table~\ref{table:Summary} and Supplemental Table~\TDemo\\

\noalign{\smallskip}
Age strata and ageing rates & Record age strata and the rate at which individuals age from one stratum to the next & Supplemental Table~\TDemo\\

\noalign{\smallskip}
Interventions & Record type, timing, effective coverage, duration, and outcomes for interventions. If multiple scenarios are reported, record representative results, i.e., record results achieved under base-case assumptions. & Tables~\ref{table:Interventions1}-\ref{table:Interventions2} and Table~\TInt\\

\noalign{\smallskip}
Calibration data & Record location, type, age stratification, time period, frequency, and original references for data used in RSV DTM calibration & Supplemental Table~\TCal\\

\noalign{\smallskip}
Parameter values & Record value and original references (if available) for common RSV DTM parameters. & Supplemental Tables~\TPar\\

\noalign{\smallskip}
Results & Record major results and findings of RSV DTMs. & Supplemental Table~\TRes\\

\noalign{\smallskip}\hline
\end{tabular}
\end{table}

\subsection{Search results}
\label{S:SearchResults}

There were $64$, $1,303$, $196$, and $1,350$ entries retrieved from Web of Science, Scopus, Embase, and PubMed searches, respectively. All searches were performed on December 01, 2020. Following removal of duplicates, titles and abstracts of the $2,672$ remaining entries were reviewed. Application of inclusion/exclusion criteria resulted in the exclusion of $2,626$ entries. Full-text manuscripts for the remaining 47 entries were retrieved and reviewed. Application of inclusion/exclusion criteria resulted in the exclusion of seven manuscripts \cite{CapistranMorelesLara2009,GuerreroFloresOsunaVargasDeLeon2019,JajarmiEtAl2020,JodarEtAl2008,ReisEtAl2019,VillanuevaOllerEtAl2013,ZhangLiuTeng2012}. The remaining 40 full-text manuscripts were included in this literature review; two manuscripts were otherwise identified and included \cite{GoldsteinEtAl2017,NugrahaNuraini2017}.

\section{RSV DTM structures}
\label{S:Structure}

\subsection{RSV disease state structure}
\label{S:DiseaseState}

\begin{figure*}
  \includegraphics[width=1\textwidth]{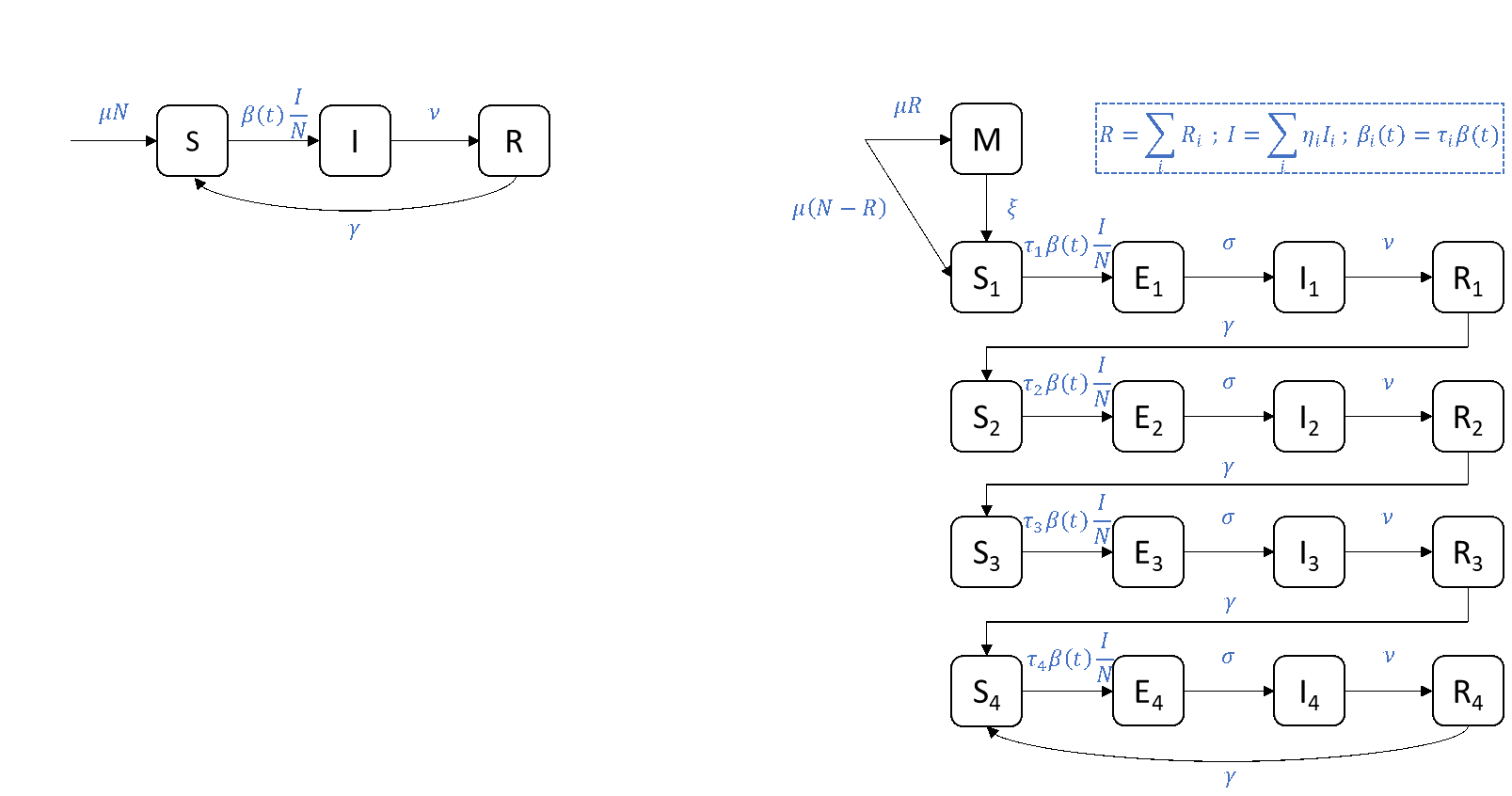}
\caption{Disease state structure for (left) $SIRS$ and (right) $M$-$SEIRS4$ RSV DTMs. Deaths, which occur from all compartments at a rate equal to the birth rate $\mu$, are omitted for clarity. The total population ($N$) is constant. The transmission term $\beta(t)$ is a one year periodic function.}
\label{figure:WWM}       
\end{figure*}

The dominant paradigm for disease state structure of RSV DTMs is established in the seminal manuscript of Weber, Weber, and Milligan \cite{WeberWeberMilligan2001}. Specifically, two disease state structures are considered: a simple susceptible-infectious-recovered-susceptible ($SIRS$) disease state structure, and a more complex $M$-$SEIRS4$ model structure (see Figure~\ref{figure:WWM} and below for definition).

The $SIRS$ model partitions individuals into three compartments: susceptible ($S$), infectious ($I$), and recovered ($R$). Infants are born into the susceptible compartment at birth rate $\mu$ and all compartments are subject to natural death at a rate equal to the birth rate, i.e., a constant population is assumed. There are three remaining transitions between compartments: susceptible individuals become infectious through contact with infectious individuals, infectious individuals recover at rate $\nu$ with full temporary immunity to reinfection, and recovered individuals become susceptible at rate $\gamma$ as full temporary immunity wanes. To account for the periodic nature of RSV epidemics, infection is assumed to occur as a result of mass action homogeneous mixing between susceptible and infectious individuals at a periodic time-varying rate proportional to
\begin{equation}
	\label{eq:beta}
	\beta(t) = b_0 \left( 1+ b_1 \cos(2\pi t-\phi) \right) \ ,
\end{equation}
where parameters $b_0$, $b_1$, and $\phi$ represent the average transmission rate, the relative amplitude of seasonal fluctuations in the transmission rate, and the phase shift of the transmission rate, respectively. In other words, new infections occur at the rate $\beta(t)SI/N$, where $N$ is the total population. 

The $M$-$SEIRS4$ model structure represents a refinement of the $SIRS$ structure based on several additional assumptions on the natural history of RSV. First, it assumes that repeated reinfection with RSV results in increasing levels of permanent partial immunity to reinfection. In contrast with full immunity, partial immunity admits reinfection (albeit at a reduced rate). In other words, a susceptible individual with zero, one, two, or three or more previous RSV infections are subdivided into compartments $S_1$, $S_2$, $S_3$, and $S_4$, respectively. Individuals in these compartments become infected at rates proportional to $\tau_1\beta(t)$, $\tau_2 \beta(t)$, $\tau_3 \beta(t)$, and $\tau_4 \beta(t)$, respectively, where $\tau_i$ are the relative susceptibilities of the different susceptible compartments. Second, as with the $SIRS$ model, infection is assumed to occur as a result of mass action homogeneous mixing between susceptible and infectious individuals, where the infectiousness of an individual with $i-1$ prior RSV infections is accounted for by multiplication with the relative infectiousness parameter $\eta_i$. In other words, new infections of individuals with $i-1$ prior RSV infections occur at the rate $\beta(t) \tau_i S_i \sum_j \eta_j I_j/N $. Third, a latency period (i.e., compartment $E_i$) is assumed where individuals are infected but not yet infectious, from which infectiousness emerges at rate $\sigma$. Finally, it is assumed that individuals are born either susceptible ($S_1$) or with full temporary immunity to RSV infection due to transfer of natural maternal antibodies ($M$). Individuals born with natural maternal antibodies are said to have natural maternal immunity (NMI) and enter compartment $M$ at a rate proportional to the fraction of recovered individuals in the population. Natural maternal immunity wanes at rate $\xi$ and results in individuals becoming susceptible. A summary of parameter and compartment definitions for the $SIR$ and $M$-$SEIRS4$ models is provided in Table~\ref{table:ParamSummary}.  

\begin{table}
\caption{Common disease states and parameters of RSV DTMs}
\label{table:ParamSummary}       
\begin{tabular}{p{3.5cm}p{7cm}}
\hline\noalign{\smallskip} 

& \textbf{Description} \\
\noalign{\smallskip}\hline\hline\noalign{\smallskip}

\textbf{Disease compartments} \\

$M$ & Natural maternal immunity\\

$S$ & Susceptible\\

$E$ & Exposed\\

$I$ & Infectious\\

$R$ &  Recovered\\

\noalign{\smallskip}\hline\noalign{\smallskip}

\textbf{Parameters}\\

$\mu$ & Birth/death rate\\

$\xi$ & Natural maternal immunity waning rate \\

$b_0$ & Average transmision rate\\

$b_1$ & Relative amplitude of seasonal fluctuations in the transmission rate\\

$\phi$ & Phase shift of the transmission rate\\

$\tau$ & Relative susceptibility to RSV infection\\

$\eta$ & Relative infectiousness while infectious with RSV\\

$\sigma$ & Rate of emergence of infectiousness\\

$\nu$ & Recovery rate\\

$\gamma$ & Immunity waning rate\\

\noalign{\smallskip}\hline
\end{tabular}
\end{table}

The $SIRS$ and $M$-$SEIRS4$ models described above are antecedent to the majority of identified RSV DTMs and admit the $(M)$-$XXXXn$ notation for similar disease state structures. If present, the prefix $M$- is used to indicate that some infants are born with NMI; otherwise, all infants are born without NMI. The body $XXXX$ is used to indicate the compartments for the progression of RSV infection. The suffix $n$ indicates the number of levels of partial immunity to RSV that are granted due to repeat reinfections. 

Whereas disease progression for the RSV DTMs that conform to the $(M)$-$XXXXn$ notation follow a simple linear pattern, disease progression in non-standard RSV DTMs are complicated by one or more of the following elements: (a) separate compartments for RSV groups A and B 
\cite{KombeEtAl2019,WhiteEtAl2005}, 
(b) multiple types of infectious compartments 
\cite{HodgsonEtAl2020,KinyanjuiEtAl2020,KombeEtAl2019,MahikulEtAl2019,PanNgumEtAl2017,YaminEtAl2016}, 
(c) waning of partial immunity to reinfection by RSV for susceptible individuals 
\cite{KinyanjuiEtAl2020,MahikulEtAl2019,PanNgumEtAl2017,WhiteEtAl2007,WhiteEtAl2005}, 
or (d) multiple nested dynamic transmission models \cite{ArenasGonzalezJodar2008,WhiteEtAl2007}.

A summary of models by their disease structure is provided in Table~\ref{table:Summary}. For completeness, a brief summary of differences in the implementation of NMI between RSV DTMs is provided in Supplementary Materials 1: Appendix~\ANMI.

\newpage
\begin{longtable}{p{1cm}p{1cm}p{2cm}p{2cm}p{2cm}p{2cm}p{4cm}}
\caption{Summary of RSV DTM structures} \label{table:Summary} \\

\hline\noalign{\smallskip} 
\textbf{Year} & \textbf{Refer-ence} & \textbf{Modelling approach} & \textbf{Country} & \textbf{Age \newline stratifica-tion} & \textbf{Demo-graphic model} & \textbf{Intervention}\\
\noalign{\smallskip}\hline\hline\noalign{\smallskip}
\endfirsthead

\multicolumn{7}{c}{{\bfseries \tablename\ \thetable{} -- continued from previous page}} \\
\hline\noalign{\smallskip} 
\textbf{Year} & \textbf{Refer-ence} & \textbf{Modelling approach} & \textbf{Country} & \textbf{Age \newline stratifica-tion} & \textbf{Demo-graphic model} & \textbf{Intervention}\\
\noalign{\smallskip}\hline\hline\noalign{\smallskip}
\endhead

\hline \multicolumn{7}{|r|}{{Continued on next page}} \\ \hline
\multicolumn{7}{p{16cm}}{\footnotesize{N/A - not applicable; ODE - ordinary differential equation; $\Delta$E - discrete difference equation;  SDE - stochastic differential equation; S$\Delta$E - stochastic difference equation; ABM - agent-based model; FDE - fractional differential equation.}}
\endfoot

\hline
\multicolumn{7}{p{16cm}}{\footnotesize{N/A - not applicable; ODE - ordinary differential equation; $\Delta$E - discrete difference equation;  SDE - stochastic differential equation; S$\Delta$E - stochastic difference equation; ABM - agent-based model; FDE - fractional differential equation.}}
\endlastfoot

\noalign{\smallskip} \multicolumn{7}{l}{\textbf{$SIRS$ models}} \\ 

\noalign{\smallskip}
2001 & \cite{WeberWeberMilligan2001} 
	& ODE
		& Finland, \newline The Gambia, \newline Singapore, \newline USA
			& N/A
				& N/A
					& N/A\\

\noalign{\smallskip}
2009 & \cite{ArenasGonzalezParraMorano2009} 
	& SDE
		& Spain
			& N/A
				& N/A
					& N/A\\

\noalign{\smallskip}
2010 & \cite{AcedoEtAl2010} 
	& ODE
		& Spain
			& Present
				& Present
					& Vaccination at birth\\

\noalign{\smallskip}
2010 & \cite{AcedoMoranoDiezDomingo2010} 
	& ODE
		& Spain
			& Present
				& Present
					& Vaccination of infants\\

\noalign{\smallskip}
2010 & \cite{ArenasGonzalezParraJodar2010} 
	& ODE
		& Spain
			& N/A
				& N/A
					& N/A\\
					
\noalign{\smallskip}
2011 & \cite{PoncianoCapistran2011} 
	& ODE
		& Finland, \newline The Gambia
			& N/A
				& N/A
					& N/A\\
					
\noalign{\smallskip}
2013 & \cite{ArandaLozanoGonzalezParraQuerales2013}
	& ODE
		& Colombia
			& N/A
				& N/A
					& N/A\\
					
\noalign{\smallskip}
2014 & \cite{CorberanValletSantonja2014}
	& S$\Delta$E
		& Spain
			& N/A
				& N/A
					& Vaccination of infants\\

\noalign{\smallskip}
2017 & \cite{NugrahaNuraini2017} 
	& ODE 
		& USA
			& N/A
				& N/A
					& Vaccination at birth,\newline Public awareness campaign\\

\noalign{\smallskip}
2017 & \cite{JornetSanzEtAl2017} 
	& S$\Delta$E 
		& Spain
			& N/A
				& N/A
					& N/A \\
					
\noalign{\smallskip}
2017 & \cite{SmithHoganMercer2017} 
	& ODE 
		& N/A
			& N/A
				& N/A
					& Maternal vaccination,\newline Vaccination (all ages)\\
					
\noalign{\smallskip}
2018 & \cite{RosaTorres2018a} 
	& ODE 
		& USA
			& N/A
				& N/A
					& Other treatment\\
					
\noalign{\smallskip}
2018 & \cite{RosaTorres2018b} 
	& FDE 
		& USA
			& N/A
				& N/A
					& Other treatment\\

\hline \noalign{\smallskip} \multicolumn{7}{l}{\textbf{$SIRSn$ models}} \\

\noalign{\smallskip}
2015 & \cite{MorrisEtAl2015} 
	& ODE 
		& N/A
			& N/A
				& N/A
					& N/A\\
					
\hline \noalign{\smallskip} \multicolumn{7}{l}{\textbf{$M$-$SIRSn$ models}} \\

\noalign{\smallskip}
2015 & \cite{KinyanjuiEtAl2015} 
	& ODE 
		& Kenya
			& Present
				& Present
					& Maternal vaccination,\newline Vaccination of infants,\newline Vaccination of school-aged children\\

\noalign{\smallskip}
2015 & \cite{PolettiEtAl2015} 
	& ABM
		& Kenya
			& Present
				& Present
					& Vaccination ($\leq10$-month-olds)\\

\noalign{\smallskip}
2017 & \cite{PanNgumEtAl2017} 
	& ODE 
		& Kenya
			& Present
				& Present
					& Maternal vaccination,\newline Vaccination of infants\\

\noalign{\smallskip}
2020 & \cite{BrandEtAl2020} 
	& ODE 
		& Kenya
			& Present
				& Present
					& Maternal vaccination,\newline Vaccination of households with newborns\\

\noalign{\smallskip}
2020 & \cite{KinyanjuiEtAl2020} 
	& ODE 
		& United Kingdom
			& Present
				& Present
					& Vaccination of infants\\
		
\hline \noalign{\smallskip} \multicolumn{7}{l}{\textbf{$SEIRS$ models}} \\				

\noalign{\smallskip}
2016 & \cite{Paynter2016} 
	& ODE 
		& Philippines
			& N/A
				& N/A
					& N/A\\
					
\noalign{\smallskip}
2018 & \cite{RosaTorres2018a} 
	& ODE 
		& USA
			& N/A
				& N/A
					& Other treatment\\
					
\noalign{\smallskip}
2018 & \cite{RosaTorres2018b} 
	& FDE 
		& USA
			& N/A
				& N/A
					& Other treatment\\

\hline \noalign{\smallskip} \multicolumn{7}{l}{\textbf{$SEIRSn$ models}} \\	

\noalign{\smallskip}
2014 & \cite{MooreEtAl2014} 
	& ODE 
		& Australia
			& Present
				& Present
					& N/A\\

\noalign{\smallskip}
2014 & \cite{PaynterEtAl2014} 
	& ODE 
		& Philippines
			& N/A
				& N/A
					& N/A\\
		
\noalign{\smallskip}
2016 & \cite{HoganEtAl2016} 
	& ODE 
		& Australia
			& Present
				& Present
					& N/A\\
					
2019 & \cite{ArguedasSantanaCibrianVelascoHernandez2019}
	& ODE 
		& Mexico
			& Present
				& Present
					& N/A\\	
					
\hline \noalign{\smallskip} \multicolumn{7}{l}{\textbf{$M$-$SEIRSn$ models}} \\	
		
\noalign{\smallskip}
2001 & \cite{WeberWeberMilligan2001} 
	& ODE
		& Finland, \newline The Gambia, \newline Singapore, \newline USA
			& N/A
				& N/A
					& N/A\\

\noalign{\smallskip}
2017 & \cite{HoganEtAl2017} 
	& ODE 
		& Australia
			& Present
				& Present
					& Maternal vaccination\\
			
\noalign{\smallskip}
2020 & \cite{CampbellGeardHogan2020} 
	& ABM 
		& Australia
			& Present
				& Present
					& Maternal vaccination\\

\hline \noalign{\smallskip} \multicolumn{7}{l}{\textbf{$SIS$ models}} \\	

\noalign{\smallskip}
2011 & \cite{MwambiEtAl2011} 
	& S$\Delta$E 
		& Kenya
			& N/A
				& N/A
					& N/A\\
					
\hline \noalign{\smallskip} \multicolumn{7}{l}{\textbf{$M$-$SIS4$ models}} \\	

\noalign{\smallskip}
2015 & \cite{PitzerEtAl2015} 
	& ODE 
		& USA
			& Present
				& Present
					& N/A\\

\hline \noalign{\smallskip} \multicolumn{7}{l}{\textbf{$SIR$ models}} \\	

2016 & \cite{ReisShaman2016} 
	& ODE 
		& USA
			& N/A
				& N/A
					& N/A\\

2018 & \cite{GoldsteinEtAl2017} 
	& $\Delta$E 
		& USA
			& Present
				& Present
					& Vaccination (various ages)\\

2018 & \cite{ReisShaman2018} 
	& ODE 
		& USA
			& N/A
				& N/A
					& N/A\\

2019 & \cite{BakerEtAl2019} 
	& S$\Delta$E 
		& Mexico,\newline USA
			& N/A
				& N/A
					& N/A\\

2020 & \cite{SeroussiLevyYomTov2020} 
	& ODE 
		& USA
			& N/A
				& N/A
					& N/A\\

2020 & \cite{vanBovenEtAl2020} 
	& ODE 
		& The Netherlands
			& Present
				& N/A
					& Maternal vaccination,\newline Vaccination of infants\\

\hline \noalign{\smallskip} \multicolumn{7}{l}{\textbf{$SEIR$ models}} \\	

2011 & \cite{LeecasterEtAl2011} 
	& ODE 
		& USA
			& Present
				& Present
					& N/A\\

\hline \noalign{\smallskip} \multicolumn{7}{l}{\textbf{Other models}} \\	

2005 & \cite{WhiteEtAl2005} 
	& ODE 
		& Finland,\newline United Kingdom
			& N/A
				& N/A
					& N/A\\
					
2007 & \cite{WhiteEtAl2007} 
	& ODE 
		& Brazil,\newline Finland,\newline The Gambia, \newline Singapore,\newline Spain,\newline United Kingdom,\newline USA
			& N/A
				& N/A
					& N/A\\

2008 & \cite{ArenasGonzalezJodar2008} 
	& ODE 
		& Brazil,\newline Spain
			& N/A
				& N/A
					& N/A\\

\noalign{\smallskip}
2016 & \cite{YaminEtAl2016} 
	& $\Delta$E 
		& USA
			& Present
				& Present
					& Vaccination (various ages)\\

\noalign{\smallskip}
2017 & \cite{PanNgumEtAl2017} 
	& ODE 
		& Kenya
			& Present
				& Present
					& Maternal vaccination,\newline Vaccination of infants\\

\noalign{\smallskip}
2019 & \cite{KombeEtAl2019} 
	& ABM 
		& Kenya
			& Present
				& Present
					& N/A\\

\noalign{\smallskip}
2019 & \cite{MahikulEtAl2019} 
	& ODE 
		& Thailand
			& Present
				& Present
					& N/A\\

\noalign{\smallskip}
2020 & \cite{HodgsonEtAl2020} 
	& ODE 
		& United Kingdom
			& Present
				& N/A
					& Maternal vaccination,\newline Vaccination of infants,\newline Vaccination of children,\newline Vaccination of older adults,\newline Monoclonal antibody immunoprophylaxis\\

\noalign{\smallskip}
2020 & \cite{KinyanjuiEtAl2020} 
	& ODE 
		& United Kingdom
			& Present
				& Present
					& Vaccination of infants\\
																						
\end{longtable}

\subsection{Demographic model structure}
\label{S:Demographic}

Demographic model structure is principally incorporated through stratification of the population by age. With the exception of three agent-based models \cite{CampbellGeardHogan2020,KombeEtAl2019,PolettiEtAl2015}, and two models restricted to $< 2$-year-olds \cite{HoganEtAl2016,Paynter2016}, age stratification uses a finer resolution for young children ($< 5$-year-olds) and a coarser resolution for adolescents, adults, and older adults 
\cite{AcedoEtAl2010,AcedoMoranoDiezDomingo2010,ArguedasSantanaCibrianVelascoHernandez2019,BrandEtAl2020,%
GoldsteinEtAl2017,HodgsonEtAl2020,HoganEtAl2017,HoganEtAl2016,KinyanjuiEtAl2015,KinyanjuiEtAl2020,KombeEtAl2019,%
LeecasterEtAl2011,MahikulEtAl2019,MooreEtAl2014,PanNgumEtAl2017,PaynterEtAl2014,PitzerEtAl2015,PolettiEtAl2015,%
vanBovenEtAl2020,YaminEtAl2016}. 
The majority of transitions between age strata are implemented at a rate proportional to the inverse of the width of the age strata of origin 
\cite{ArguedasSantanaCibrianVelascoHernandez2019,BrandEtAl2020,HodgsonEtAl2020,HoganEtAl2017,HoganEtAl2016,%
KinyanjuiEtAl2015,KinyanjuiEtAl2020,LeecasterEtAl2011,MooreEtAl2014,PanNgumEtAl2017,PitzerEtAl2015,vanBovenEtAl2020}. 
Alternatively, some RSV DTMs implement more complicated ageing schemes to maintain realistic age structures 
\cite{AcedoEtAl2010,AcedoMoranoDiezDomingo2010,KinyanjuiEtAl2020,YaminEtAl2016}. 
A few models provide additional demographic structure through organization of the simulation population by household 
\cite{BrandEtAl2020,CampbellGeardHogan2020,KombeEtAl2019,MahikulEtAl2019}, 
household and primary school \cite{PolettiEtAl2015}, and geography \cite{SeroussiLevyYomTov2020}. A more detailed discussion of demographic structure is presented in Supplemental Materials~1: Appendix~\ADemo.

\subsection{Interventions}
\label{S:Interventions}

The most common intervention considered is vaccination or monoclonal immunoprophylaxis that induces full temporary immunity to RSV infection 
\cite{AcedoEtAl2010,AcedoMoranoDiezDomingo2010,BrandEtAl2020,GoldsteinEtAl2017,HodgsonEtAl2020,JornetSanzEtAl2017,KinyanjuiEtAl2015,NugrahaNuraini2017,vanBovenEtAl2020}; 
however, vaccination inducing partial temporary immunity to RSV infection 
\cite{HoganEtAl2017,KinyanjuiEtAl2020,PanNgumEtAl2017,SmithHoganMercer2017,YaminEtAl2016}, 
public awareness campaigns \cite{NugrahaNuraini2017}, and treatment \cite{RosaTorres2018b,RosaTorres2018a}, are also considered. Interventions are generally assumed to occur uniformly throughout the year, however, exceptions include models that assume vaccination occurs at arbitrary time points throughout the year \cite{SmithHoganMercer2017}, seasonally according to the pattern observed in influenza vaccination \cite{YaminEtAl2016}, at enrollment of primary school \cite{PolettiEtAl2015}, and prior to the RSV season \cite{GoldsteinEtAl2017,HodgsonEtAl2020}. Target populations are typically newborn and infants 
\cite{AcedoEtAl2010,AcedoMoranoDiezDomingo2010,HodgsonEtAl2020,HoganEtAl2017,JornetSanzEtAl2017,KinyanjuiEtAl2015,%
KinyanjuiEtAl2020,NugrahaNuraini2017,PanNgumEtAl2017,PolettiEtAl2015,SmithHoganMercer2017,vanBovenEtAl2020},  
or young children \cite{HodgsonEtAl2020,PolettiEtAl2015}, however, vaccination of all age strata are also considered \cite{GoldsteinEtAl2017,HodgsonEtAl2020,YaminEtAl2016}. Maternal vaccination is also sometimes considered 
\cite{BrandEtAl2020,CampbellGeardHogan2020,HodgsonEtAl2020,HoganEtAl2017,PanNgumEtAl2017,PolettiEtAl2015,%
SmithHoganMercer2017,vanBovenEtAl2020}, 
however, the effect of maternal vaccination on the mother is frequently omitted \cite{HoganEtAl2017,PanNgumEtAl2017,SmithHoganMercer2017,vanBovenEtAl2020}. Additional details on model interventions are included in Supplemental Materials~1: Appendix~\AIntervention. 

\subsection{Modelling techniques}
\label{S:Techniques}

Whereas the majority of models are implemented as ordinary differential equation (ODE)-type models 
\cite{AcedoEtAl2010,ArandaLozanoGonzalezParraQuerales2013,ArenasGonzalezJodar2008,ArenasGonzalezParraJodar2010,%
ArguedasSantanaCibrianVelascoHernandez2019,BrandEtAl2020,HodgsonEtAl2020,HoganEtAl2017,HoganEtAl2016,%
KinyanjuiEtAl2015,KinyanjuiEtAl2020,LeecasterEtAl2011,MahikulEtAl2019,MooreEtAl2014,MorrisEtAl2015,NugrahaNuraini2017,%
PanNgumEtAl2017,Paynter2016,PaynterEtAl2014,PitzerEtAl2015,PoncianoCapistran2011,ReisShaman2016,ReisShaman2018,%
RosaTorres2018a,SeroussiLevyYomTov2020,SmithHoganMercer2017,vanBovenEtAl2020,WeberWeberMilligan2001,%
WhiteEtAl2007,WhiteEtAl2005}, 
there have also been RSV DTMs implemented as stochastic differential equation (SDE) models \cite{ArenasGonzalezParraMorano2009}, discrete difference equation ($\Delta$E) models \cite{GoldsteinEtAl2017,YaminEtAl2016}, stochastic difference equation models (S$\Delta$E) \cite{BakerEtAl2019,CorberanValletSantonja2014,JornetSanzEtAl2017,MwambiEtAl2011}, agent-based models (ABMs) \cite{AcedoMoranoDiezDomingo2010,CampbellGeardHogan2020,KombeEtAl2019,PolettiEtAl2015}, and fractional differential equation (FDE) models \cite{RosaTorres2018b}, see Table~\ref{table:Summary}. 

ODE and $\Delta$E models use a deterministic modelling approach that is specified in continuous and discrete time, respectively. These approaches are relatively well understood, can be solved relatively quickly (i.e., with low computational cost), and are easily adapted to many dynamic systems. These models perform best at predicting the average outcome under the assumption of a large well-mixed population.

SDE and S$\Delta$E models are extensions of ODE and $\Delta$E models, respectively, that incorporate random effects. For example, two SDE models were developed in order to study interseason variance in RSV epidemics \cite{ArenasGonzalezParraMorano2009}. Similarly, S$\Delta$E models have been developed to study infection dynamics when only small numbers of infectious individuals are present \cite{CorberanValletSantonja2014,JornetSanzEtAl2017}. Alternatively, whereas other models require the specification of a functional form for the time-varying transmission rate $\beta(t)$, two S$\Delta$E models have been developed to estimate (a) the transmission rate $\beta(t)$ as a function of time \cite{MwambiEtAl2011}, and (b) both the number of susceptible individuals and the transmission rate $\beta(t)$ as a function of time \cite{BakerEtAl2019}.  

ABMs are characterized by their specification of rules for the behaviors of individual agents. ABMs admit a granular demographic structure that is generally not considered in standard ODE models, e.g., they are capable of organizing the population into households \cite{CampbellGeardHogan2020,KombeEtAl2019}, or households and primary schools \cite{PolettiEtAl2015}.

Finally, FDE models represent a new non-local modelling approach that introduces a form of “memory” \cite{DuWangHu2013}, in which the future evolution of a FDE model simultaneously depends upon its present state and its past states. An initial FDE model has been proposed \cite{RosaTorres2018b}, however, it is unclear what advantages, if any, exist that would justify the additional complexity of FDE models over the alternatives proposed above. 

\section{Parameterization and calibration}
\label{S:PandC}

The RSV DTMs summarized above have been calibrated to diverse data sets collected from more than a dozen countries, i.e., 
Australia \cite{CampbellGeardHogan2020,HoganEtAl2017,HoganEtAl2016,MooreEtAl2014}, 
Brazil \cite{WhiteEtAl2007}, 
Colombia \cite{ArandaLozanoGonzalezParraQuerales2013}, 
Finland \cite{PoncianoCapistran2011,WeberWeberMilligan2001,WhiteEtAl2007,WhiteEtAl2005}, 
The Gambia\cite{PoncianoCapistran2011,WeberWeberMilligan2001,WhiteEtAl2007}, 
Kenya \cite{BrandEtAl2020,KinyanjuiEtAl2015,KombeEtAl2019,PanNgumEtAl2017,PolettiEtAl2015}, 
Mexico \cite{ArguedasSantanaCibrianVelascoHernandez2019,BakerEtAl2019}, 
The Netherlands \cite{vanBovenEtAl2020}, 
Philippines \cite{PaynterEtAl2014}, 
Singapore \cite{WeberWeberMilligan2001,WhiteEtAl2007}, 
Spain \cite{AcedoEtAl2010,AcedoMoranoDiezDomingo2010,ArenasGonzalezParraJodar2010,%
ArenasGonzalezParraMorano2009,CorberanValletSantonja2014,JornetSanzEtAl2017,WhiteEtAl2007}, 
Thailand \cite{MahikulEtAl2019}, 
the United Kingdom \cite{HodgsonEtAl2020,KinyanjuiEtAl2020,WhiteEtAl2007,WhiteEtAl2005}, 
and the United States \cite{BakerEtAl2019,GoldsteinEtAl2017,LeecasterEtAl2011,NugrahaNuraini2017,PitzerEtAl2015,%
ReisShaman2016,ReisShaman2018,RosaTorres2018b,RosaTorres2018a,SeroussiLevyYomTov2020,WeberWeberMilligan2001,%
WhiteEtAl2007,YaminEtAl2016}. 
These data mostly consist of RSV detected in inpatient settings only, i.e., hospitalizations, or in inpatient and outpatient settings. One model uses Google searches for the term “RSV” as a proxy for the number of RSV infections \cite{SeroussiLevyYomTov2020}. Data have been gathered for infants ($< 1$-year-olds) 
\cite{AcedoEtAl2010,AcedoMoranoDiezDomingo2010,CampbellGeardHogan2020,CorberanValletSantonja2014,%
JornetSanzEtAl2017,WhiteEtAl2005}, 
toddlers ($< 2$-year-olds) \cite{HoganEtAl2017,HoganEtAl2016,MooreEtAl2014,PaynterEtAl2014,PoncianoCapistran2011,%
WeberWeberMilligan2001,WhiteEtAl2007}, 
young children ($< 5$-year-olds) \cite{ArandaLozanoGonzalezParraQuerales2013,ArenasGonzalezParraJodar2010,%
ArenasGonzalezParraMorano2009,HodgsonEtAl2020,KinyanjuiEtAl2015,KinyanjuiEtAl2020,PanNgumEtAl2017,PolettiEtAl2015,%
WhiteEtAl2007}, 
children \cite{LeecasterEtAl2011,HodgsonEtAl2020,NugrahaNuraini2017,PoncianoCapistran2011,%
WeberWeberMilligan2001,WhiteEtAl2007,WhiteEtAl2005}, 
and the entire population \cite{ArguedasSantanaCibrianVelascoHernandez2019,BakerEtAl2019,BrandEtAl2020,%
GoldsteinEtAl2017,HodgsonEtAl2020,KombeEtAl2019,MahikulEtAl2019,PitzerEtAl2015,ReisShaman2016,ReisShaman2018,%
SeroussiLevyYomTov2020,vanBovenEtAl2020,WeberWeberMilligan2001,WhiteEtAl2007,YaminEtAl2016}. 
Frequency of measurements are daily \cite{LeecasterEtAl2011}, biweekly \cite{KombeEtAl2019}, 
weekly \cite{AcedoEtAl2010,AcedoMoranoDiezDomingo2010,ArandaLozanoGonzalezParraQuerales2013,%
ArguedasSantanaCibrianVelascoHernandez2019,BrandEtAl2020,CorberanValletSantonja2014,HodgsonEtAl2020,HoganEtAl2016,%
JornetSanzEtAl2017,KinyanjuiEtAl2020,MooreEtAl2014,PitzerEtAl2015,PolettiEtAl2015,PoncianoCapistran2011,ReisShaman2016,%
ReisShaman2018,SeroussiLevyYomTov2020,vanBovenEtAl2020,WeberWeberMilligan2001,WhiteEtAl2007,WhiteEtAl2005,%
YaminEtAl2016}, 
monthly \cite{ArenasGonzalezParraJodar2010,ArenasGonzalezParraMorano2009,HoganEtAl2017,KinyanjuiEtAl2015,MooreEtAl2014,%
MahikulEtAl2019,NugrahaNuraini2017,PanNgumEtAl2017,PoncianoCapistran2011,RosaTorres2018b,RosaTorres2018a,%
WeberWeberMilligan2001,WhiteEtAl2007}, 
or annually \cite{GoldsteinEtAl2017,WhiteEtAl2005}. For additional details, see Supplemental Materials~1: Appendix~\ACalibration

This review has compiled values for comparison of common model parameters determined through calibration or literature search. Comparison of parameter values has value in not only populating future RSV DTMs, but also in identifying uncertainty in aspects of the natural history of RSV that may require further research to resolve. Results for four common parameters are summarized: NMI waning rate ($\xi$), relative susceptibility to RSV infection ($\tau$), recover rate ($\nu$), and immunity waning rate ($\gamma$). A comprehensive summary of common parameter values is provided in Supplementary Materials~1: Appendix~\AParameter.

\subsection{Natural immunity waning rate ($\xi$)}
\label{S:xi}

Seven models estimate NMI waning rate from literature values. For the five most recent models the NMI waning rate lies within the range $2.7-4.1$ per year, equivalent to a duration of $90-134$ days 
\cite{CampbellGeardHogan2020,HodgsonEtAl2020,PitzerEtAl2015,PolettiEtAl2015,YaminEtAl2016}; 
for the remaining two models the NMI waning rate is 13.0 per year, equivalent to a duration of 28 days 
\cite{ArenasGonzalezParraMorano2009,WeberWeberMilligan2001}. 
In contrast, the calibration of six models produces estimates of the NMI waning rate in the range $5.2-49.6$ per year, equivalent to a duration of $7-70$ days 
\cite{BrandEtAl2020,KinyanjuiEtAl2015,KinyanjuiEtAl2020,PanNgumEtAl2017}. 
Comparison of these values indicate some uncertainty exists in the NMI waning rate which may require additional research to resolve.

\subsection{Relative susceptibility to RSV infection ($\tau$)}
\label{S:tau}

Five models estimate relative susceptibility of individuals with at least one previous RSV infection ($\tau_1$) to be in the range $0.45-0.77$, when measured with respect to the reference susceptibility of RSV naïve individuals ($\tau_0 = 1$) 
\cite{BrandEtAl2020,KinyanjuiEtAl2020,MahikulEtAl2019,MorrisEtAl2015,PaynterEtAl2014}. 
In contrast, calibration of two models produces estimates in the range $0.68-0.88$ \cite{PolettiEtAl2015,WhiteEtAl2007}. These values are largely consistent and give insight into the approximate range for relative susceptibility of individuals previously infected with RSV.

\subsection{Recovery rate ($\nu$)}
\label{S:nu}

Using literature values, twenty three papers estimate the recovery rate to be in the range $33.2-46.8$ per year, equivalent to a duration of $8-11$ days 
\cite{AcedoEtAl2010,AcedoMoranoDiezDomingo2010,ArandaLozanoGonzalezParraQuerales2013,ArenasGonzalezJodar2008,%
ArenasGonzalezParraJodar2010,ArenasGonzalezParraMorano2009,CampbellGeardHogan2020,CorberanValletSantonja2014,%
GoldsteinEtAl2017,HoganEtAl2017,HoganEtAl2016,JornetSanzEtAl2017,LeecasterEtAl2011,MooreEtAl2014,NugrahaNuraini2017,%
PolettiEtAl2015,PoncianoCapistran2011,RosaTorres2018b,RosaTorres2018a,SmithHoganMercer2017,WeberWeberMilligan2001,%
WhiteEtAl2007,WhiteEtAl2005}. 
In contrast, calibration of two models produces estimates in the range $57.0-70.2$ per year, equivalent to a duration of $5-6$ days \cite{ReisShaman2016,ReisShaman2018}. As with the NMI waning rate, the discrepancy between literature and calibration estimates may indicate some uncertainty in the recovery rate; however, it is noted that the models that estimate recovery rate through calibration employ an $SIR$ model structure to model each season separately, i.e., they depart from the standard $(M)$-$XXXXn$ disease structure typically employed in RSV DTMs.

\subsection{Immunity waning rate ($\gamma$)}
\label{S:gamma}

Nineteen papers use literature values to estimate a range for the immunity waning rate: $1.8-2.0$ per year, equivalent to a duration of $183-203$ days 
\cite{AcedoEtAl2010,AcedoMoranoDiezDomingo2010,ArandaLozanoGonzalezParraQuerales2013,%
ArenasGonzalezParraJodar2010,ArenasGonzalezParraMorano2009,BrandEtAl2020,CorberanValletSantonja2014,%
JornetSanzEtAl2017,KinyanjuiEtAl2015,KinyanjuiEtAl2020,MorrisEtAl2015,NugrahaNuraini2017,PanNgumEtAl2017,%
PoncianoCapistran2011,RosaTorres2018b,RosaTorres2018a,SmithHoganMercer2017,WeberWeberMilligan2001,YaminEtAl2016}. 
One model uses literature to estimate a rate of 5.8 per year, equivalent to a duration of 63 days \cite{PaynterEtAl2014}. One model uses literature to estimate a rate of 1.0 per year, equivalent to a duration of 359 days \cite{HodgsonEtAl2020}. In contrast, calibration of three models produces estimates of the immunity waning rate in the range $1.6-2.1$ per year, equivalent to $171-230$ days \cite{HoganEtAl2016,MooreEtAl2014,PolettiEtAl2015}. These values are largely consistent and give insight into an approximate range for immunity waning rate.

\section{Modelling results}
\label{S:Results}

In this section we summarize some important modelling results of the RSV DTMs reviewed above. For additional details see Supplementary Materials~1: Appendix~\AResults.

\subsection{General modelling results}
\label{S:GeneralResults}

The $SIRS$ and $M$-$SEIRS4$ RSV DTMs introduced above \cite{WeberWeberMilligan2001} (see Figure 2) establish a disease state structure that informs, directly or indirectly, the disease state structure of most subsequent RSV DTMs. A sensitivity analysis performed on the $SIRS$ ODE model by varying initial conditions, birth rate ($\mu$), and average transmission parameter ($b_0$) finds that the model is least sensitive to uncertainty in initial conditions and most sensitive to uncertainty in average transmission parameter \cite{ArenasGonzalezParraJodar2010}. Consistent results are reported for an $SIRS$ SDE models \cite{ArenasGonzalezParraMorano2009}.

Hogan and colleagues performed an analysis of an age-stratified $SEIRS$ ODE model that provides some insight into the behavior of models implementing the $(M)$-$XXXXn$ disease state structure \cite{HoganEtAl2016}. The simple $SEIRS$ model was able to reproduce the diverse periodic behaviors observed in RSV epidemics: an annual pattern of repeating peaks, a biennial pattern of repeating high followed by low peaks where peaks occur at the same time each year, and a biennial pattern of high followed by low peaks where high peaks occur earlier in the year than low peaks. Roughly speaking, annual peaks result when the duration of immunity ($1/\gamma$) is short, the former biennial pattern results when the average transmission coefficient ($b_1$) is large, and the latter biennial pattern for intermediate values of the birth rate ($\mu$).

Additional insight into the $(M)$-$XXXXn$ disease structure is provided by comparing a system of eight standard nested models (including, e.g., $SIS$, $SIR$, and $SIRS$ model structures, among others) on their ability to reproduce RSV epidemic data \cite{WhiteEtAl2007}. The most parsimonious model with the best fit was a model with partial permanent immunity following an initial infection with RSV. These results are consistent with the majority of models that implement the $(M)$-$XXXXn$ disease structure, a structure with increasing levels of partial permanent immunity resulting from repeated RSV infections. Additional evidence supporting the inclusion of partial permanent immunity following initial infection with RSV is provided by Morris and colleagues \cite{MorrisEtAl2015}, who conclude that the $SIRS2$ model is better able to capture sensitivity of RSV epidemics to birth rate than the $SIRS$ model.

\subsection{Results for RSV interventions}
\label{S:InterventionsResults}

Direct comparison of modelling results is complicated by several factors. First, different parameter ranges are considered for vaccine effective coverage (the product of vaccine coverage and effectiveness), and duration of protection. Second, models differ by the mechanism of protection (e.g., full temporary immunity versus partial temporary immunity). Third, outcomes are measured with respect to different populations. Comparison of modelling results are, therefore, qualitative in nature.

Seven models report reduction in hospitalizations or infections due to maternal vaccination 
\cite{BrandEtAl2020,CampbellGeardHogan2020,HodgsonEtAl2020,HoganEtAl2017,PanNgumEtAl2017,PolettiEtAl2015,%
vanBovenEtAl2020}. 
These models exhibit four different mechanisms of protection:  full temporary immunity provided to both mother and infant \cite{BrandEtAl2020,HodgsonEtAl2020,PolettiEtAl2015}, full temporary immunity provided to mother and partial temporary immunity provided to infant \cite{CampbellGeardHogan2020},  full temporary immunity provided to infant only \cite{vanBovenEtAl2020}, and partial temporary immunity provided to infant only \cite{HoganEtAl2017,PanNgumEtAl2017}, see Table~\ref{table:Interventions1} for representative results. For effective coverage of $35\%-60\%$ and duration of protection of $3-6$ months, the reduction in hospitalizations of infants ($< 1$-year-olds) is approximately $6\%-20\%$ and the reduction in infections of infants ($< 1$-year-olds) is approximately $17\%-26\%$. 

\begin{table}
\caption{Reduction in hospitalization or infection due to maternal vaccination}
\label{table:Interventions1}       
\begin{tabular}{p{1cm}p{1cm}p{1.5cm}p{1.5cm}p{3cm}p{1.5cm}}
\hline\noalign{\smallskip}
\textbf{Year} & \textbf{Refer-ence} & \textbf{Effective coverage (\%)} & \textbf{Duration (months)} & \textbf{Reference \newline population \newline(age in months)} & \textbf{Percent reduction (\%)}  \\
\noalign{\smallskip}\hline\hline\noalign{\smallskip}

\multicolumn{6}{l}{\textbf{Newborns granted PTI}}\\

\noalign{\smallskip}
2017 
	& \cite{HoganEtAl2017}
		& 40
			& 6
				& $0-2$ & $6-37$\textsuperscript{a} \\
 	& 	& 	& 	& $3-5$ & $30-46$\textsuperscript{a} \\
 	&	&	&  	& $6-11$ & 0\textsuperscript{a}\\
 	&	&	& 3	& $0-2$	& 25\textsuperscript{a}\\
 	&	&	& 	& $3-5$	& 0\textsuperscript{a}\\
 	&	&	& 	& $6-11$ & 0\textsuperscript{a}\\

\noalign{\smallskip}
2017 
	& \cite{PanNgumEtAl2017}
		& 35
			& 3
				& $<12$ & $7-15$\textsuperscript{a} \\

\hline\noalign{\smallskip}				
\multicolumn{6}{l}{\textbf{Newborns granted FTI}}\\

\noalign{\smallskip}	
2020
	& \cite{vanBovenEtAl2020}
		& 50
			& 6
				& $<12$ & 26\textsuperscript{b} \\

\hline\noalign{\smallskip}				
\multicolumn{6}{l}{\textbf{Newborns granted PTI; mothers granted FTI}}\\

\noalign{\smallskip}	
2020
	& \cite{CampbellGeardHogan2020}
		& N/A\textsuperscript{c}
			& 3
				& $<3$ & 17\textsuperscript{b} \\
	&	&	&	& $3-6$ & 5\textsuperscript{b} \\

\hline\noalign{\smallskip}				
\multicolumn{6}{l}{\textbf{Newborns and mothers granted FTI}}\\

\noalign{\smallskip}	
2015
	& \cite{PolettiEtAl2015}
		& 60
			& 6
				& $<12$ & 17\textsuperscript{b} \\

\noalign{\smallskip}	
2020
	& \cite{BrandEtAl2020}
		& 50
			& 3
				& $<60$ & 19\textsuperscript{a} \\
				
\noalign{\smallskip}	
2020
	& \cite{HodgsonEtAl2020}
		& 32
			& 4
				& All ages & 9\textsuperscript{a} \\

\noalign{\smallskip}\hline
\multicolumn{6}{l}{\footnotesize{PTI - partial temporary immunity; FTI - full temporary immunity}}\\
\multicolumn{6}{l}{\footnotesize{\textsuperscript{a} Percent reduction in hospitalizations}}\\
\multicolumn{6}{l}{\footnotesize{\textsuperscript{b} Percent reduction in infections}}\\
\multicolumn{6}{l}{\footnotesize{\textsuperscript{c} Coverage was 70\%; effectiveness was not specified.}}
\end{tabular}
\end{table}

Seven models report reduction in hospitalizations or infections due to infant vaccination or monoclonal immunoprophylaxis 
\cite{HodgsonEtAl2020,JornetSanzEtAl2017,KinyanjuiEtAl2015,KinyanjuiEtAl2020,PanNgumEtAl2017,PolettiEtAl2015,%
vanBovenEtAl2020}. 
Analogous to maternal vaccination, these models exhibit two different mechanisms of protection: partial temporary immunity \cite{KinyanjuiEtAl2020,PanNgumEtAl2017}, and full temporary immunity 
\cite{HodgsonEtAl2020,KinyanjuiEtAl2015,JornetSanzEtAl2017,vanBovenEtAl2020}, 
see Table~\ref{table:Interventions2} for representative results. For effective coverage of $80\%-90\%$ and duration of protection of $6-12$ months, the reduction of hospitalizations of infants ($< 1$-year-olds) is approximately $50\%-90\%$ and the reduction in infections of infants ($< 1$-year-olds) is approximately $30\%-35\%$. 

\begin{table}
\caption{Reduction in hospitalization due to infant vaccination}
\label{table:Interventions2}       
\begin{tabular}{p{1cm}p{1cm}p{1.5cm}p{1.5cm}p{3cm}p{1.5cm}}
\hline\noalign{\smallskip}
\textbf{Year} & \textbf{Refer-ence} & \textbf{Effective coverage (\%)} & \textbf{Duration (months)} & \textbf{Reference \newline population \newline(age in months)} & \textbf{Percent reduction (\%)}  \\
\noalign{\smallskip}\hline\hline\noalign{\smallskip}

\multicolumn{6}{l}{\textbf{Infants granted PTI}}\\

\noalign{\smallskip}
2017 
	& \cite{PanNgumEtAl2017}
		& 90
			& 12
				& $<12$ & $58-89$\textsuperscript{a} \\

\noalign{\smallskip}
2020 
	& \cite{KinyanjuiEtAl2020}
		& 90
			& 12
				& $<12$ & $55-56$\textsuperscript{a} \\

\hline\noalign{\smallskip}				
\multicolumn{6}{l}{\textbf{Infants granted FTI}}\\

\noalign{\smallskip}	
2015
	& \cite{KinyanjuiEtAl2015}
		& 80
			& 6
				& $<6$ & $51-88$\textsuperscript{a} \\
				
\noalign{\smallskip}	
2015
	& \cite{PolettiEtAl2015}
		& 80
			& 6
				& $<12$ & $35$\textsuperscript{b} \\

\noalign{\smallskip}	
2017
	& \cite{JornetSanzEtAl2017}
		& 80
			& 6
				& $<24$ & $81$\textsuperscript{a} \\

\noalign{\smallskip}	
2020
	& \cite{HodgsonEtAl2020}
		& 75
			& 12
				& All ages & 7\textsuperscript{a} \\
	&	& 63\textsuperscript{c} 
			& 8\textsuperscript{c}
				& All ages\textsuperscript{c} & 8\textsuperscript{a,c}\\

\noalign{\smallskip}	
2020
	& \cite{vanBovenEtAl2020}
		& 50
			& 55
				& $<12$ & $30$\textsuperscript{b} \\

\noalign{\smallskip}\hline
\multicolumn{6}{l}{\footnotesize{PTI - partial temporary immunity; FTI - full temporary immunity}}\\
\multicolumn{6}{l}{\footnotesize{\textsuperscript{a} Percent reduction in hospitalizations}}\\
\multicolumn{6}{l}{\footnotesize{\textsuperscript{b} Percent reduction in infections}}\\
\multicolumn{6}{p{10cm}}{\footnotesize{\textsuperscript{c} Long-acting immunoprophylaxis administered to all infants at birth (if born in-season) or at the beginning of the season (if born out-of-season).}}
\end{tabular}
\end{table}

A hybrid approach is studied by Brand and colleagues \cite{BrandEtAl2020}, in which maternal vaccination is combined with vaccination of the entire household at birth. Maternal vaccination is assumed to provide newborns with an additional 75 days of protection (for a total of 96 days of protection), vaccination of household members is assumed to provide six months of protection, and protection for both forms of vaccination is assumed to take the form of full temporary immunity. Under these assumptions, an effective coverage of 75\% of birth households results in a 50\% reduction in RSV hospitalizations of under 5-year-olds.

Three models compare vaccination of multiple age groups \cite{GoldsteinEtAl2017,HodgsonEtAl2020,YaminEtAl2016}. These results of these three studies are consistent, i.e., it is found that vaccination of under 5-year-olds is the most efficient strategy for averting RSV infection \cite{HodgsonEtAl2020,YaminEtAl2016}, and vaccination of $3-6$-year-olds at the beginning of the RSV season results in the greatest reduction in the initial effective reproduction number \cite{GoldsteinEtAl2017}.

Finally, two models provide a cost-effectiveness analysis for a hypothetical vaccine for infants in Valencia, Spain \cite{AcedoEtAl2010,AcedoMoranoDiezDomingo2010}. These models include hospitalization cost, vaccination cost, and parent/caregiver loss of productivity, and find that cost savings are possible when average parent/caregiver loss of productivity exceeds three days per infant infected with RSV. One model provides a cost-effectiveness analysis for palivizumab and three hypothetical products: a maternal vaccine, a vaccine, and a long-acting monoclonal immunoprophylaxis \cite{HodgsonEtAl2020}. This model includes costs for administering the vaccine or immunoprophylaxis, hospitalization, and general practice visits, and calculates the maximum cost-effective purchase price for various comparators, see Supplemental Materials~1: Appendix~\AResults\ for additional details. 

\subsection{Seasonal drivers}
\label{S:DriversResults}

Most models presented in this review assume that there exists some periodic forcing of RSV epidemic dynamics, e.g., see Equation~\ref{eq:beta}. Three models explore potential drivers of this seasonal forcing in detail \cite{BakerEtAl2019,PaynterEtAl2014,PitzerEtAl2015}. In the Philippines the peak in RSV transmission is found to precede the peak in RSV detections by $49-67$ days, and nutritional status and rainfall are identified as two potential drivers of RSV epidemic dynamics \cite{PaynterEtAl2014}. In the United States correlation is observed between estimated model parameters and climatic variables of temperature, vapor pressure, precipitation, and potential evapotranspiration \cite{PitzerEtAl2015}. For example, the relative amplitude of seasonal fluctuations in the transmission rate ($b_1$) and the phase shift of the transmission rate ($\phi$) were found to be negatively correlated with mean precipitation and mean vapor pressure, and positively correlated with the amplitude and timing of potential evapotranspiration . Similarly, a more recent modelling paper covering both the United States and Mexico finds an inverse relationship between humidity and log transmission and a positive linear relationship between rainfall and transmission rate \cite{BakerEtAl2019}. Additionally, one papers estimates the seasonal transmission rate $\beta(t)$ as a function of time for an RSV epidemic in Kilifi, Kenya, and finds two peaks in transmission (May, and January/February) \cite{MwambiEtAl2011}.

\subsection{Forecasting RSV epidemics}
\label{S:ForecastingResults}

Four models were developed with application to forecasting RSV epidemic dynamics \cite{LeecasterEtAl2011,ReisShaman2016,ReisShaman2018,SeroussiLevyYomTov2020}. First, the average transmission coefficient ($b_0$) and epidemic start time estimated from consecutive seasons are found to covary, and are potentially predictive of epidemic size \cite{LeecasterEtAl2011}. Second, an $SIR$ model calibrated to data in real time is developed as a forecasting model that, four weeks prior to the peak in RSV detections, predicts the magnitude of the peak in RSV detections within 25\% approximately 70\% of the time \cite{ReisShaman2016,ReisShaman2018}. Finally, a multicompartment $SIR$ model for the United States is capable of predicting infection rates and timing of infection peaks with high accuracy in each state for the current season using the first seven weeks of RSV data and parameters estimated from the previous year’s data \cite{SeroussiLevyYomTov2020}.

\section{Research gaps and future steps}
\label{S:Gaps}

The diversity of RSV DTMs and their applications admits numerous opportunities for improvement in our understanding of RSV epidemic dynamics. Below, four areas with significant potential for future work are discussed: alignment of RSV DTMs with immunoprophylactic profiles, understanding sensitivity of results to model structure, evaluation of cost-effectiveness through health economic analysis, and investigating seasonal drivers of RSV epidemics.

\subsection{Alignment with immunoprophylactic profiles}
\label{S:AlignIP}

Because there are currently no available vaccines or immunoprophylactic interventions for RSV, the RSV DTMs identified above are necessarily limited to implementing hypothetical products. As the profiles of the products under development become more well defined, further alignment with RSV DTMs will be possible. Further stratification of the model to include additional sub-populations of interest may be necessary. For example, none of the current RSV DTMs include compartments for high-risk infants, i.e., very premature infants, infants with CHD, or infants with CLD. In particular, stratification by gestational age may be important when evaluating maternal vaccination strategies, since transfer of maternal antibodies for preterm infants is expected to be incomplete \cite{RainischEtAl2020}. Analogously, evaluations of interventions targeted at older children and adults should also consider inclusion of high-risk subpopulations, i.e., older adults, institutionalized adults, and immunocompromised adults. 

\subsection{Sensitivity of results to model structure}
\label{S:Sensitivity}

Only a few efforts have been made to determine the sensitivity of modelling results to model structure. Indeed, a full understanding of the sensitivity of modelling results to model structure may not be possible due to the complexity observed in RSV DTMs. As models are developed and aligned to specific product profiles, comparison between model predictions may become a practical strategy to validate models and to achieve insights into the sensitivity of results to model structure. 

\subsection{Health economic analysis}
\label{S:HECON}

As the development of RSV immunoprophylactic products continues to advance, health economic analyses will become increasingly important tools for informing public health decision making. Whereas health economic analyses that employ static modelling approaches will continue to play an important role, given the highly contagious nature of RSV, we anticipate increasing demand for analyses that are better suited to address indirect effects or herd immunity effects. In other words, we anticipate increasing demand for health economic analyses based on dynamic modelling approaches, i.e., DTMs. Studies that describe a cost-effectiveness analysis based on an RSV DTMs, a subset of all manuscripts that include an RSV DTM, are identified in only three manuscripts \cite{AcedoEtAl2010,AcedoMoranoDiezDomingo2010,HodgsonEtAl2020}. There is significant potential for additional RSV DTMs that evaluate costs related to RSV infection in multiple countries and settings. Additionally, there is potential for future work comparing and contrasting health economic analyses that employ static modelling approaches to those that employ RSV DTMs.

\subsection{Seasonal drivers of RSV}
\label{S:Seasonality}

Finally, the mechanisms by which potential drivers affect epidemic dynamics have not been described in detail. In other words, seasonality in RSV DTMs is typically incorporated through an exogenous forcing term, e.g., see Equation~\ref{eq:beta}, that has an arbitrary functional form. Additional research into the sensitivity of RSV DTM results to the functional form of the seasonal forcing term may lead to improvements in how seasonality is included in RSV DTMs. A better understanding of how to include seasonal drivers of RSV epidemics into RSV DTMs may allow for more accurate models, better predictions of changing patterns in RSV epidemics (e.g., due to climate change), and identification of more efficient intervention strategies.  

\section{Discussion}
\label{S:Discussion}

\subsection{Stengths and limitations}
\label{S:SandL}

This literature review, the first literature review of RSV DTMs, provides a comprehensive summary of RSV DTMs. Broad search terms were used and over 2,600 titles and abstracts were reviewed in order to identify 38 full-text manuscripts for inclusion (two additional manuscripts were otherwise identified). The manuscripts included in this review represent a diversity of RSV DTMs and admits a broad overview of RSV DTMs provided along multiple dimensions (e.g., disease state structure, underlying demographic model structure, interventions included, calibration method and data, and modelling techniques applied), and perspectives (e.g., analytical/theoretical, epidemiologic, health economic). Furthermore, the Supplementary Materials that accompany this review are a potentially valuable resource. For example, the Supplementary Materials include (but are not limited to) a detailed description of all data sets used in calibration of RSV DTMs, a detailed description of common parameter values used in parameterization of RSV DTMs, and a detailed description of interventions included in RSV DTMs; where applicable, the original references to these additional data have also been provided.

This review is subject to several limitations. First, because risk of bias and quality for RSV DTMs (and DTMs in general) is context dependent, we do not provide an assessment of the risk of bias or quality of the included RSV DTMs. Indeed, the risk of bias or quality in an RSV DTM depends not only on model structure, input parameters, calibration data, modelling technique, et cetera, but also on the modelling objectives. For example, a simple $SIR$ model may be appropriate to the forecasting of RSV epidemics, but it is completely incapable of assessing the impact of maternal vaccination on infant hospitalizations. The lack of a critical appraisal of included studies is not unusual for literature reviews \cite{MunnEtAl2018}, and a full risk of bias and quality assessment is left as future work, e.g., as part of a future systematic review with a specific research question. Second, this review remains subject to evidence selection bias. Specifically, although this review conducted a very broad search in multiple databases, the choice to include only RSV DTMs presented in published manuscripts admits the risk of publication bias. Finally, this review was entirely completed by a single author. Although all steps in this review were completed in duplicate, we may still expect a higher error rate in screening manuscripts and data abstraction than if this review were conducted by multiple authors working independently and aggregating their results. 

\subsection{Conclusions}
\label{S:Conclusions}

The numerous vaccines and immunoprophylactic interventions currently under development for prevention of RSV infection have the potential to significantly reduce the burden of RSV in infants in the near future. Mathematical modelling provides a means to better understand the natural history of RSV, to forecast severity of RSV epidemics mid-season, to predict long-term changes in patterns of RSV epidemics, and to evaluate the effectiveness of proposed vaccine and immunoprophylactic interventions. This review has provided an overview of existing RSV DTMs that includes disease state structures, demographic model structure, intervention strategies, and modelling techniques. In both the main text and the Supplementary Materials, a list of RSV epidemic data sources and values of common parameters determined through literature and calibration has been compiled. This work provides a strong foundation for future modelling of RSV epidemics and interventions. Research gaps and areas for future potential work have also been identified. In particular, it is anticipated that RSV DTMs, combined with economic cost-effectiveness evaluations, will play a significant role in shaping decision making in the development and implementation of vaccination and immunoprophylaxis programs.

\newpage
\bibliographystyle{spmpsci}      

\begin{thebibliography}{10}
\providecommand{\url}[1]{{#1}}
\providecommand{\urlprefix}{URL }
\expandafter\ifx\csname urlstyle\endcsname\relax
  \providecommand{\doi}[1]{DOI~\discretionary{}{}{}#1}\else
  \providecommand{\doi}{DOI~\discretionary{}{}{}\begingroup
  \urlstyle{rm}\Url}\fi

\bibitem{BradyEtAl2014}
{Updated Guidance for Palivizumab Prophylaxis Among Infants and Young Children
  at Increased Risk of Hospitalization for Respiratory Syncytial Virus
  Infection}.
\newblock Pediatrics \textbf{134}(2), e620--e638 (2014).
\newblock \doi{10.1542/peds.2014-1666}

\bibitem{AcedoEtAl2010}
Acedo, L., D\'{i}ez-Domingo, J., Mora\~{n}o, J.A., Villanueva, R.J.:
  Mathematical modelling of respiratory syncytial virus ({RSV}): vaccination
  strategies and budget applications.
\newblock Epidemiol Infect \textbf{138}(6), 853–860 (2010).
\newblock \doi{10.1017/S0950268809991373}

\bibitem{AcedoMoranoDiezDomingo2010}
Acedo, L., Moraño, J.A., Díez-Domingo, J.: {Cost analysis of a vaccination
  strategy for respiratory syncytial virus (RSV) in a network model}.
\newblock Math Comput Model \textbf{52}(7), 1016 -- 1022 (2010).
\newblock \doi{10.1016/j.mcm.2010.02.041}

\bibitem{ArandaLozanoGonzalezParraQuerales2013}
Aranda-Lozano, D., Gonz\'alez-Parra, G., Querales, J.: {Modelamiento de la
  transmisi\'on del Virus Respiratorio Sincitial (VRS) en ni\~nos menores de
  cinco a\~nos}.
\newblock Revista de Salud P\'ublica \textbf{15}(4), 689--700 (2013)

\bibitem{ArenasGonzalezJodar2008}
Arenas, A., González, G., Jódar, L.: {Existence of periodic solutions in a
  model of respiratory syncytial virus RSV}.
\newblock J Math Anal Appl \textbf{344}(2), 969--980 (2008).
\newblock \doi{10.1016/j.jmaa.2008.03.049}

\bibitem{ArenasGonzalezParraJodar2010}
Arenas, A., González-Parra, G., Jódar, L.: {Randomness in a mathematical
  model for the transmission of respiratory syncytial virus (RSV)}.
\newblock Math Comput Simulat \textbf{80}(5), 971--981 (2010).
\newblock \doi{10.1016/j.matcom.2009.12.001}

\bibitem{ArenasGonzalezParraMorano2009}
Arenas, A., González-Parra, G., Moraño, J.A.: {Stochastic modeling of the
  transmission of respiratory syncytial virus (RSV) in the region of Valencia,
  Spain}.
\newblock Biosystems \textbf{96}(3), 206--212 (2009).
\newblock \doi{10.1016/j.biosystems.2009.01.007}

\bibitem{ArguedasSantanaCibrianVelascoHernandez2019}
Arguedas, Y., Santana-Cibrian, M., Velasco-Hern\'andez, J.: Transmission
  dynamics of acute respiratory diseases in a population structured by age.
\newblock Math Biosci Eng \textbf{16}, 7477 (2019).
\newblock \doi{10.3934/mbe.2019375}

\bibitem{BakerEtAl2019}
Baker, R., Mahmud, A., Wagner, C., {et al.}: Epidemic dynamics of respiratory
  syncytial virus in current and future climates.
\newblock Nat Commun \textbf{10}, 5512 (2019).
\newblock \doi{10.1038/s41467-019-13562-y}

\bibitem{BloomEtAl2013}
Bloom-Feshbach, K., Alonso, W., Charu, V., {et al}: {Latitudinal Variations in
  Seasonal Activity of Influenza and Respiratory Syncytial Virus (RSV): A
  Global Comparative Review}.
\newblock PLOS ONE \textbf{8}(2), 1--12 (2013).
\newblock \doi{10.1371/journal.pone.0054445}

\bibitem{BrandEtAl2020}
Brand, S., Munywoki, P., Walumbe, D., {et al}: Reducing {RSV} hospitalisation
  in a lower-income country by vaccinating mothers-to-be and their households.
\newblock eLife \textbf{9}, e47003 (2020).
\newblock \doi{10.7554/eLife.47003}

\bibitem{CampbellGeardHogan2020}
Campbell, P., Geard, N., Hogan, A.: {Modelling the household-level impact of a
  maternal respiratory syncytial virus (RSV) vaccine in a high-income setting}.
\newblock BMC Med \textbf{18}, 319 (2020).
\newblock \doi{10.1186/s12916-020-01783-8}

\bibitem{CapistranMorelesLara2009}
Capistr\'{a}n, M., Moreles, M., Lara, B.: {Parameter Estimation of Some
  Epidemic Models. The Case of Recurrent Epidemics Caused by Respiratory
  Syncytial Virus}.
\newblock B Math Biol \textbf{71}, 1890--1901 (2009).
\newblock \doi{10.1007/s11538-009-9429-3}

\bibitem{ChubbJacobsen2010}
Chubb, M., Jacobsen, K.: Mathematical modeling and the epidemiological research
  process.
\newblock Eur J Epidemiol \textbf{25}, 13--19 (2010).
\newblock \doi{10.1007/s10654-009-9397-9}

\bibitem{WoS2020}
{Clarivate Analytics}: {Web of Science} ({2020}).
\newblock \urlprefix\url{https://webofknowledge.com/}.
\newblock Accessed 12/01/2020.

\bibitem{CorberanValletSantonja2014}
Corberán-Vallet, A., Santonja, F.: {A Bayesian SIRS model for the analysis of
  respiratory syncytial virus in the region of Valencia, Spain}.
\newblock Biometrical J \textbf{56}(5), 808--818 (2014).
\newblock \doi{10.1002/bimj.201300194}

\bibitem{DuWangHu2013}
Du, M., Wang, Z., Hu, H.: Measuring memory with the order of fractional
  derivative.
\newblock Sci Rep \textbf{3}, 3431 (2013).
\newblock \doi{10.1038/srep03431}

\bibitem{Elsevier2020b}
Elsevier: Embase (2020).
\newblock \urlprefix\url{https://embase.com}.
\newblock Accessed 12/01/2020.

\bibitem{Elsevier2020a}
Elsevier: Scopus (2020).
\newblock \urlprefix\url{https://scopus.com}.
\newblock Accessed 12/01/2020.

\bibitem{FalseyEtAl2005}
Falsey, A., Hennessey, P., Formica, M., {et al}: {Respiratory Syncytial Virus
  Infection in Elderly and High-Risk Adults}.
\newblock New Engl J Med \textbf{352}(17), 1749--1759 (2005).
\newblock \doi{10.1056/NEJMoa043951}

\bibitem{GlezenEtAl1986}
Glezen, W., Taber, L., Frank, A., Kasel, J.: {Risk of Primary Infection and
  Reinfection With Respiratory Syncytial Virus}.
\newblock Am J Dis Child \textbf{140}(6), 543--546 (1986).
\newblock \doi{10.1001/archpedi.1986.02140200053026}

\bibitem{GoldsteinEtAl2017}
Goldstein, E., Nguyen, H., Liu, P., {et al}: {On the Relative Role of Different
  Age Groups During Epidemics Associated With Respiratory Syncytial Virus}.
\newblock J Infect Dis \textbf{217}(2), 238--244 (2018).
\newblock \doi{10.1093/infdis/jix575}

\bibitem{GuerreroFloresOsunaVargasDeLeon2019}
Guerrero-Flores, S., Osuna, O., {Vargas-De-Le{\'o}n}, C.: {Periodic solutions
  for seasonal SIQRS models with nonlinear infection terms}.
\newblock Electron J Differ Eq  (2019)

\bibitem{GutfraindEtAl2015}
Gutfraind, A., Galvani, A., Meyers, L.: {Efficacy and Optimization of
  Palivizumab Injection Regimens Against Respiratory Syncytial Virus
  Infection}.
\newblock JAMA Pediatr \textbf{169}(4), 341--348 (2015).
\newblock \doi{10.1001/jamapediatrics.2014.3804}

\bibitem{HallEtAl2001}
Hall, C., Long, C., Schnabel, K.: {Respiratory Syncytial Virus Infections in
  Previously Healthy Working Adults}.
\newblock Clin Infect Dis \textbf{33}(6), 792--796 (2001).
\newblock \doi{10.1086/322657}

\bibitem{HallEtAl2013}
Hall, C., Weinberg, G., Blumkin, A., {et al}: {Respiratory Syncytial
  Virus{\textendash}Associated Hospitalizations Among Children Less Than 24
  Months of Age}.
\newblock Pediatrics \textbf{132}(2), e341--e348 (2013).
\newblock \doi{10.1542/peds.2013-0303}

\bibitem{HallEtAl2009}
Hall, C., Weinberg, G., Iwane, M., {et al}: {The Burden of Respiratory
  Syncytial Virus Infection in Young Children}.
\newblock New Engl J Med \textbf{360}(6), 588--598 (2009).
\newblock \doi{10.1056/NEJMoa0804877}

\bibitem{HendersonEtAl1979}
Henderson, F., Collier, A., Clyde, W., Denny, F.: {Respiratory-Syncytial-Virus
  Infections, Reinfections and Immunity}.
\newblock New Engl J Med \textbf{300}(10), 530--534 (1979).
\newblock \doi{10.1056/NEJM197903083001004}.
\newblock PMID: 763253

\bibitem{HigginsEtAl2016}
Higgins, D., Trujillo, C., Keech, C.: {Advances in RSV vaccine research and
  development – A global agenda}.
\newblock Vaccine \textbf{34}(26), 2870--2875 (2016).
\newblock \doi{10.1016/j.vaccine.2016.03.109}

\bibitem{HodgsonEtAl2020}
Hodgson, D., Pebody, R., Panovska-Griffiths, J., , {et al}: {Evaluating the
  next generation of RSV intervention strategies: a mathematical modelling
  study and cost-effectiveness analysis}.
\newblock BMC Med \textbf{18}, 348 (2020).
\newblock \doi{10.1186/s12916-020-01802-8}

\bibitem{HoganEtAl2017}
Hogan, A., Campbell, P., Blyth, C., {et al}: {Potential impact of a maternal
  vaccine for RSV: A mathematical modelling study}.
\newblock Vaccine \textbf{35}(45), 6172 -- 6179 (2017).
\newblock \doi{10.1016/j.vaccine.2017.09.043}

\bibitem{HoganEtAl2016}
Hogan, A., Glass, K., Moore, H., Anderssen, R.: {Exploring the dynamics of
  respiratory syncytial virus (RSV) transmission in children}.
\newblock Theor Popul Biol \textbf{110}, 78 -- 85 (2016).
\newblock \doi{10.1016/j.tpb.2016.04.003}

\bibitem{JajarmiEtAl2020}
Jajarmi, A., Yusuf, A., Baleanu, D., Inc, M.: {A new fractional HRSV model and
  its optimal control: A non-singular operator approach}.
\newblock Physica A \textbf{547}, 123860 (2020).
\newblock \doi{10.1016/j.physa.2019.123860}

\bibitem{JodarEtAl2008}
Jódar, L., Villanueva, R., Arenas, A.: {Modeling the spread of seasonal
  epidemiological diseases: Theory and applications}.
\newblock Math Comput Model \textbf{48}(3), 548--557 (2008).
\newblock \doi{10.1016/j.mcm.2007.08.017}

\bibitem{JornetSanzEtAl2017}
Jornet-Sanz, M., Corberán-Vallet, A., Santonja, F., Villanueva, R.: {A
  Bayesian stochastic SIRS model with a vaccination strategy for the analysis
  of respiratory syncytial virus}.
\newblock SORT-Stat Oper Res T \textbf{1}(1), 159--176 (2017)

\bibitem{KinyanjuiEtAl2015}
Kinyanjui, T., House, T., Kiti, M., {et al}: {Vaccine Induced Herd Immunity for
  Control of Respiratory Syncytial Virus Disease in a Low-Income Country
  Setting}.
\newblock PLOS ONE \textbf{10}(9), 1--16 (2015).
\newblock \doi{10.1371/journal.pone.0138018}

\bibitem{KinyanjuiEtAl2020}
Kinyanjui, T., Pan-Ngum, W., Saralamba, S., {et al}: Model evaluation of target
  product profiles of an infant vaccine against respiratory syncytial virus
  ({RSV}) in a developed country setting.
\newblock Vaccine X \textbf{4}, 100055 (2020).
\newblock \doi{10.1016/j.jvacx.2020.100055}

\bibitem{KombeEtAl2019}
Kombe, I., Munywoki, P., Baguelin, M., {et al}: Model-based estimates of
  transmission of respiratory syncytial virus within households.
\newblock Epidemics \textbf{27}, 1 -- 11 (2019).
\newblock \doi{10.1016/j.epidem.2018.12.001}

\bibitem{LeecasterEtAl2011}
Leecaster, M., Gesteland, P., Greene, T., {et al}: Modeling the variations in
  pediatric respiratory syncytial virus seasonal epidemics.
\newblock BMC Infect Dis \textbf{11}, 105 (2011)

\bibitem{LiEtAl2019}
Li, Y., Reeves, R., Wang, X., {et al}: Global patterns in monthly activity of
  influenza virus, respiratory syncytial virus, parainfluenza virus, and
  metapneumovirus: a systematic analysis.
\newblock Lancet Glob Health \textbf{7}(8), e1031--e1045 (2019).
\newblock \doi{10.1016/S2214-109X(19)30264-5}

\bibitem{MahikulEtAl2019}
Mahikul, W., White, L., Poovorawan, K., {et al}: {Modeling household dynamics
  on Respiratory Syncytial Virus (RSV)}.
\newblock PLOS ONE \textbf{14}(7), 1--13 (2019).
\newblock \doi{10.1371/journal.pone.0219323}

\bibitem{MooreEtAl2014}
Moore, H., Jacoby, P., Hogan, A., {et al}: {Modelling the Seasonal Epidemics of
  Respiratory Syncytial Virus in Young Children}.
\newblock PLOS ONE \textbf{9}(6), 1--8 (2014).
\newblock \doi{10.1371/journal.pone.0100422}

\bibitem{MorrisEtAl2015}
Morris, S., Pitzer, V., Viboud, C., {et al}: {Demographic buffering: titrating
  the effects of birth rate and imperfect immunity on epidemic dynamics}.
\newblock J R Soc Interface \textbf{12}(104), 20141245 (2015).
\newblock \doi{10.1098/rsif.2014.1245}

\bibitem{MunnEtAl2018}
Munn, Z., Peters, M., Stern, C., {et al}: {Systematic review or scoping review?
  Guidance for authors when choosing between a systematic or scoping review
  approach}.
\newblock BMC Med Res Methodol \textbf{18}, 143 (2018).
\newblock \doi{10.1186/s12874-018-0611-x}

\bibitem{MwambiEtAl2011}
Mwambi, H., Ramroop, S., White, L., {et al}: {A frequentist approach to
  estimating the force of infection for a respiratory disease using repeated
  measurement data from a birth cohort}.
\newblock Stat Methods Med Res \textbf{20}(5), 551--570 (2011).
\newblock \doi{10.1177/0962280210385749}

\bibitem{NairEtAl2010}
Nair, H., Nokes, D., Gessner, B., {et al}: Global burden of acute lower
  respiratory infections due to respiratory syncytial virus in young children:
  a systematic review and meta-analysis.
\newblock Lancet \textbf{375}(9725), 1545--1555 (2010).
\newblock \doi{10.1016/S0140-6736(10)60206-1}

\bibitem{NairEtAl2013}
Nair, H., Simões, E., Rudan, I., {et al}: {Global and regional burden of
  hospital admissions for severe acute lower respiratory infections in young
  children in 2010: a systematic analysis}.
\newblock Lancet \textbf{381}(9875), 1380--1390 (2013).
\newblock \doi{10.1016/S0140-6736(12)61901-1}

\bibitem{NCBI2020}
{National Center for Biotechnology Information}: {PubMed} (2020).
\newblock \urlprefix\url{https://pubmed.ncbi.nlm.nih.gov/}.
\newblock Accessed 12/01/2020.

\bibitem{NugrahaNuraini2017}
Nugraha, E., Nuraini, N.: {Simple Vaccination and Prevention Model of
  Respiratory Syncytial Virus}.
\newblock Far East J Math Sci \textbf{102}(9), 1865--1880 (2017)

\bibitem{PanNgumEtAl2017}
Pan-Ngum, W., Kinyanjui, T., Kiti, M., {et al}: {Predicting the relative
  impacts of maternal and neonatal respiratory syncytial virus (RSV) vaccine
  target product profiles: A consensus modelling approach}.
\newblock Vaccine \textbf{35}(2), 403 -- 409 (2017).
\newblock \doi{10.1016/j.vaccine.2016.10.073}

\bibitem{PATH2020}
{PATH}: {RSV Vaccine and mAb Snapshot} (2020).
\newblock
  \urlprefix\url{https://www.path.org/resources/rsv-vaccine-and-mab-snapshot/}.
\newblock Accessed 08/25/2020.

\bibitem{Paynter2016}
Paynter, S.: {Incorporating Transmission Into Causal Models of Infectious
  Diseases for Improved Understanding of the Effect and Impact of Risk
  Factors}.
\newblock Am J Epidemiol \textbf{183}(6), 574--582 (2016).
\newblock \doi{10.1093/aje/kwv234}

\bibitem{PaynterEtAl2014}
Paynter, S., Yakob, L., Simões, E., {et al}: {Using Mathematical Transmission
  Modelling to Investigate Drivers of Respiratory Syncytial Virus Seasonality
  in Children in the Philippines}.
\newblock PLOS ONE \textbf{9}(2), 1--11 (2014).
\newblock \doi{10.1371/journal.pone.0090094}

\bibitem{PitmanEtAl2012}
Pitman, R., Fisman, D., Zaric, G., {et al}: {Dynamic Transmission Modeling: A
  Report of the ISPOR-SMDM Modeling Good Research Practices Task Force Working
  Group–5}.
\newblock Med Decis Making \textbf{32}(5), 712--721 (2012).
\newblock \doi{10.1177/0272989X12454578}.
\newblock PMID: 22990086

\bibitem{PitzerEtAl2015}
Pitzer, V., Viboud, C., Alonso, W., {et al}: {Environmental Drivers of the
  Spatiotemporal Dynamics of Respiratory Syncytial Virus in the United States}.
\newblock PLOS Pathog \textbf{11}(1), 1--14 (2015).
\newblock \doi{10.1371/journal.ppat.1004591}

\bibitem{PolettiEtAl2015}
Poletti, P., Merler, S., Ajelli, M., {et al}: Evaluating vaccination strategies
  for reducing infant respiratory syncytial virus infection in low-income
  settings.
\newblock BMC Med \textbf{13}, 49 (2015).
\newblock \doi{10.1186/s12916-015-0283-x}

\bibitem{PoncianoCapistran2011}
Ponciano, J., Capistrán, M.: {First Principles Modeling of Nonlinear Incidence
  Rates in Seasonal Epidemics}.
\newblock PLOS Comput Biol \textbf{7}(2), 1--14 (2011).
\newblock \doi{10.1371/journal.pcbi.1001079}

\bibitem{RainischEtAl2020}
Rainisch, G., Adhikari, B., Meltzer, M., Langley, G.: {Estimating the impact of
  multiple immunization products on medically-attended respiratory syncytial
  virus (RSV) infections in infants}.
\newblock Vaccine \textbf{38}(2), 251--257 (2020).
\newblock \doi{10.1016/j.vaccine.2019.10.023}

\bibitem{ReisShaman2016}
Reis, J., Shaman, J.: {Retrospective Parameter Estimation and Forecast of
  Respiratory Syncytial Virus in the United States}.
\newblock PLOS Comput Biol \textbf{12}(10), 1--15 (2016).
\newblock \doi{10.1371/journal.pcbi.1005133}

\bibitem{ReisShaman2018}
Reis, J., Shaman, J.: Simulation of four respiratory viruses and inference of
  epidemiological parameters.
\newblock Infect Dis Model \textbf{3}, 23 -- 34 (2018).
\newblock \doi{10.1016/j.idm.2018.03.006}

\bibitem{ReisEtAl2019}
Reis, J., Yamana, T., Kandula, S., Shaman, J.: {Superensemble forecast of
  respiratory syncytial virus outbreaks at national, regional, and state levels
  in the United States}.
\newblock Epidemics \textbf{26}, 1--8 (2019).
\newblock \doi{10.1016/j.epidem.2018.07.001}

\bibitem{RhaEtAl2020}
Rha, B., Curns, A., Lively, J., {et al}: Respiratory syncytial
  virus{\textendash}associated hospitalizations among young children:
  2015{\textendash}2016.
\newblock Pediatrics \textbf{146}(1) (2020).
\newblock \doi{10.1542/peds.2019-3611}

\bibitem{RosaTorres2018b}
Rosa, S., Torres, D.: {Optimal control of a fractional order epidemic model
  with application to human respiratory syncytial virus infection}.
\newblock Chaos Soliton Fract \textbf{117}, 142 -- 149 (2018).
\newblock \doi{10.1016/j.chaos.2018.10.021}

\bibitem{RosaTorres2018a}
Rosa, S., Torres, D.: {Parameter Estimation, Sensitivity Analysis and Optimal
  Control of a Periodic Epidemic Model with Application to HRSV in Florida}.
\newblock Stat Optim Inform Comput \textbf{6}, 139 -- 149 (2018).
\newblock \doi{10.19139/soic.v6i1.472/j.chaos.2018.10.021}

\bibitem{SeroussiLevyYomTov2020}
Seroussi, I., Levy, N., Yom-Tov, E.: Multi-season analysis reveals the spatial
  structure of disease spread.
\newblock Physica A \textbf{547}, 124425 (2020).
\newblock \doi{10.1016/j.physa.2020.124425}

\bibitem{SmithHoganMercer2017}
Smith, R., Hogan, A., Mercer, G.: {Unexpected Infection Spikes in a Model of
  Respiratory Syncytial Virus Vaccination}.
\newblock Vaccines \textbf{5}(2), 1--15 (2017).
\newblock \doi{10.3390/vaccines5020012}

\bibitem{vanBovenEtAl2020}
{van Boven}, M., Teirlinck, A., Meijer, A., {et al}: {Estimating Transmission
  Parameters for Respiratory Syncytial Virus and Predicting the Impact of
  Maternal and Pediatric Vaccination}.
\newblock J Infect Dis \textbf{222}(Supplement 7), S688--S694 (2020).
\newblock \doi{10.1093/infdis/jiaa424}

\bibitem{VillanuevaOllerEtAl2013}
Villanueva-Oller, J., Acedo, L., Mora\~{n}o, J., S\'{a}nchez-S\'{a}nchez, A.:
  {Epidemic Random Network Simulations in a Distributed Computing Environment}.
\newblock Abstr Appl Anal \textbf{2013}, 462801 (2013).
\newblock \doi{10.1155/2013/462801}

\bibitem{WeberWeberMilligan2001}
Weber, A., Weber, M., Milligan, P.: {Modeling epidemics caused by respiratory
  syncytial virus (RSV)}.
\newblock Math Biosci \textbf{172}(2), 95--113 (2001)

\bibitem{WhiteEtAl2007}
White, L., Mandl, J., Gomes, M., {et al}: Understanding the transmission
  dynamics of respiratory syncytial virus using multiple time series and nested
  models.
\newblock Math Biosci \textbf{209}(1), 222 -- 239 (2007).
\newblock \doi{10.1016/j.mbs.2006.08.018}

\bibitem{WhiteEtAl2005}
White, L.J., Waris, M., Cane, P.A., {et al}: {The transmission dynamics of
  groups A and B human respiratory syncytial virus (hRSV) in England \& Wales
  and Finland: seasonality and cross-protection}.
\newblock Epidemiol Infect \textbf{133}(2), 279–289 (2005).
\newblock \doi{10.1017/S0950268804003450}

\bibitem{WidmerEtAl2014}
Widmer, K., Griffin, M., Zhu, Y., {et al}: Respiratory syncytial virus- and
  human metapneumovirus-associated emergency department and hospital burden in
  adults.
\newblock Influenza Other Resp \textbf{8}(3), 347--352 (2014).
\newblock \doi{10.1111/irv.12234}

\bibitem{WidmerEtAl2012}
Widmer, K., Zhu, Y., Williams, J., {et al}: {Rates of Hospitalizations for
  Respiratory Syncytial Virus, Human Metapneumovirus, and Influenza Virus in
  Older Adults}.
\newblock J Infect Dis \textbf{206}(1), 56--62 (2012).
\newblock \doi{10.1093/infdis/jis309}

\bibitem{YaminEtAl2016}
Yamin, D., Jones, F., DeVincenzo, J., {et al}: Vaccination strategies against
  respiratory syncytial virus.
\newblock P Natl Acad Sci USA \textbf{113}(46), 13239--13244 (2016).
\newblock \doi{10.1073/pnas.1522597113}

\bibitem{ZhangLiuTeng2012}
Zhang, T., Liu, J., Ten, Z.: {Existence of positive periodic solutions of an
  SEIR model with periodic coefficients}.
\newblock Appl Math \textbf{57}, 601--616 (2012).
\newblock \doi{10.1007/s10492-012-0036-5}

\end{thebibliography}


\end{document}